\documentclass{JINST}
\usepackage{cite}

\title{A method to suppress dielectric breakdowns in liquid argon ionization detectors 
for cathode to ground distances of several millimeters}

\author{ M.~Auger, A.~Ereditato,  D.~Goeldi, S.~Janos, I.~Kreslo\thanks{Corresponding author.}~, 
M.~Luethi, C.~Rudolf~von~Rohr,  T.~Strauss, T. Tolba and M.~S.~Weber\\
\llap Albert Einstein Center for Fundamental Physics, Laboratory for High Energy Physics \\
  Universit\"{a}t Bern, Switzerland\\
E-mail: \email{igor.kreslo@lhep.unibe.ch}}

\abstract{We present a method to reach electric field intensity as high as 400 kV/cm in liquid argon for cathode-ground distances of several millimeters. This can be achieved by suppressing field emission from the cathode, overcoming limitations that we reported earlier.}


\begin{document}
\section{Introduction}

Obtaining the required electric field in Time Projection Chambers (TPCs) based on liquefied noble gases, such as argon or xenon, is related to the electrical strength of these media and is limited by the risk of electrical breakdowns in regions typically outside the drift volume. This is an important issue when dealing with the high potential values (from 100~kV to 1~MV) that are required for long drift length detectors, such as ARGONTUBE \cite{ARGONTUBE2}, MicroBooNE  \cite{uboone} or the envisioned neutrino observatories LBNO \cite{LBNO} and LBNE \cite{LBNE}. Although the drift field inside the field cage for such detectors is of the order of 1~kV/cm,  the field outside of the cage, especially in the gap between the cathode and the grounded walls of the cryostat, is substantially higher.

So far, a commonly used reference value for the dielectric strength of liquid argon
was of the order of 1~MV/cm (first measured and reported in \cite{SWAN60,SWAN61}).
In \cite{Blatter:2014wua} we have shown that the breakdown in liquid argon can occur at electric field intensities as low as 40 kV/cm at the surface of a mechanically polished stainless steel cathode, when the cathode to ground distance is of the order of a centimeter. This sets serious constraints on the design of TPC detector geometries implying relatively long distances between the field shaping cage and the cryostat walls. The outer edges radii of the field cage must also be large, in order to minimize surface fields. This leads to the requirement of a large amount of liquid outside  the sensitive volume of the detector, hence increasing its size and cost.

A suggested mechanism for the initiation of breakdown discharges invokes field emission of electrons from the metal of the cathode, enhanced by the positive argon ions concentrating near its surface. Therefore, in order to suppress the discharges, one needs to inhibit the initial field emission. In fact, the effect of reducing field emission and increasing the breakdown voltage in liquid argon by oxidation of the cathode metal surface was already reported in literature \cite{SWAN60,stefan1,stefan2}. In the following, we report on a method to achieve a significant suppression of the breakdowns by deposition of a layer of dielectric polymer on the cathode.

\section{Experimental setup and measurement results}

\begin{figure}[ht]
\centering	
\includegraphics[width=0.8\linewidth]{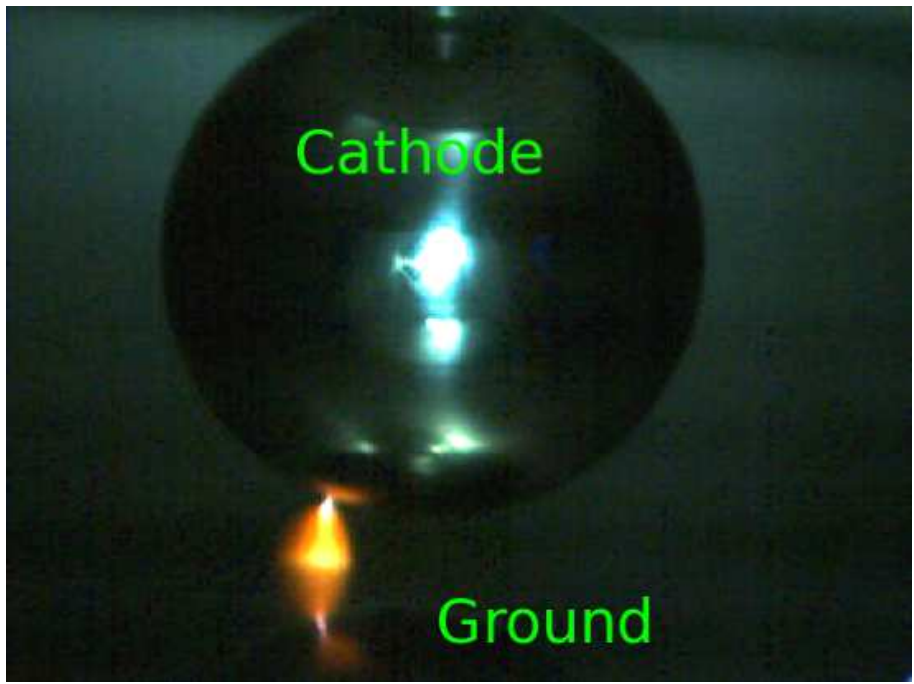}
\caption{Initial stage of a breakdown: a massive electron emission from the cathode that excites and ionizes the liquid argon is visible (bright orange cone). In the lower part a reflection of the cathode and the emission cone from the ground plane can be seen.}
\label{fig:image1}
\end{figure}

The setup we used for the study reported here is identical to the one employed and described in \cite{Blatter:2014wua}.
A spherical cathode and a plane anode, both mechanically polished to a mirror state, formed the discharge gap. 
The purity of the liquid argon after filling was estimated with a small TPC according to the method described in \cite{Badhrees:2010zz}. It was found to be of the order of 1~ppb of oxygen-equivalent impurity concentration. 
The setup was complemented by a high-speed camera to visually observe the development of discharges. The camera is capable of recording 700x400 pixel RGB images at 1250~fps.
A typical initial stage of the breakdown, as registered by the camera, is shown in Figure~\ref{fig:image1}. A massive electron emission from the cathode that excites and ionizes the liquid argon in the initial stage of the discharge is clearly visible.

We managed to largely suppress the field emission by depositing a layer of polymer onto the cathode sphere surface. The dielectric material must have high dielectric strength and has to keep a suitable elasticity at the 87K temperature of liquid argon. It must also have a low excess electron mobility compared to that of liquid argon (saturated drift velocity of about 7 mm/$\mu$s at 100 kV/cm, from \cite{review}) in order to allow the accumulation of the negative charge, that reduces the cathode surface field intensity and suppresses field emission.

A polymer that satisfies the above requirements is natural polyisoprene (latex rubber). This material can be easily deposited onto the polished metal surface from an emulsified water solution (latex milk). The reported dielectric strength of the resulting layer ranges from 1 to 2 MV/cm \cite{Skanavi}. Its dielectric constant is 2.1, close to 1.6 of liquid argon, and the resistivity is 10$^{16}$~Ohm$\cdot$cm at room temperature. At 87K the film of natural rubber retains some elasticity, so that a several hundred micrometers thick layer deposited at the surface of a few centimeter diameter steel sphere keeps its integrity and does not crack even under multiple fast cooling-down and warming-up cycles.
The mobility of excess electrons in polyisoprene has not been reported in literature so far. Existing data for similar polymers fall in the range of 10$^{-11}$ to 10$^{-4}$~cm$^2$/Vs (see $e.g.$ \cite{mobility}).

\begin{figure}
\centering	
\includegraphics[width=0.45\linewidth]{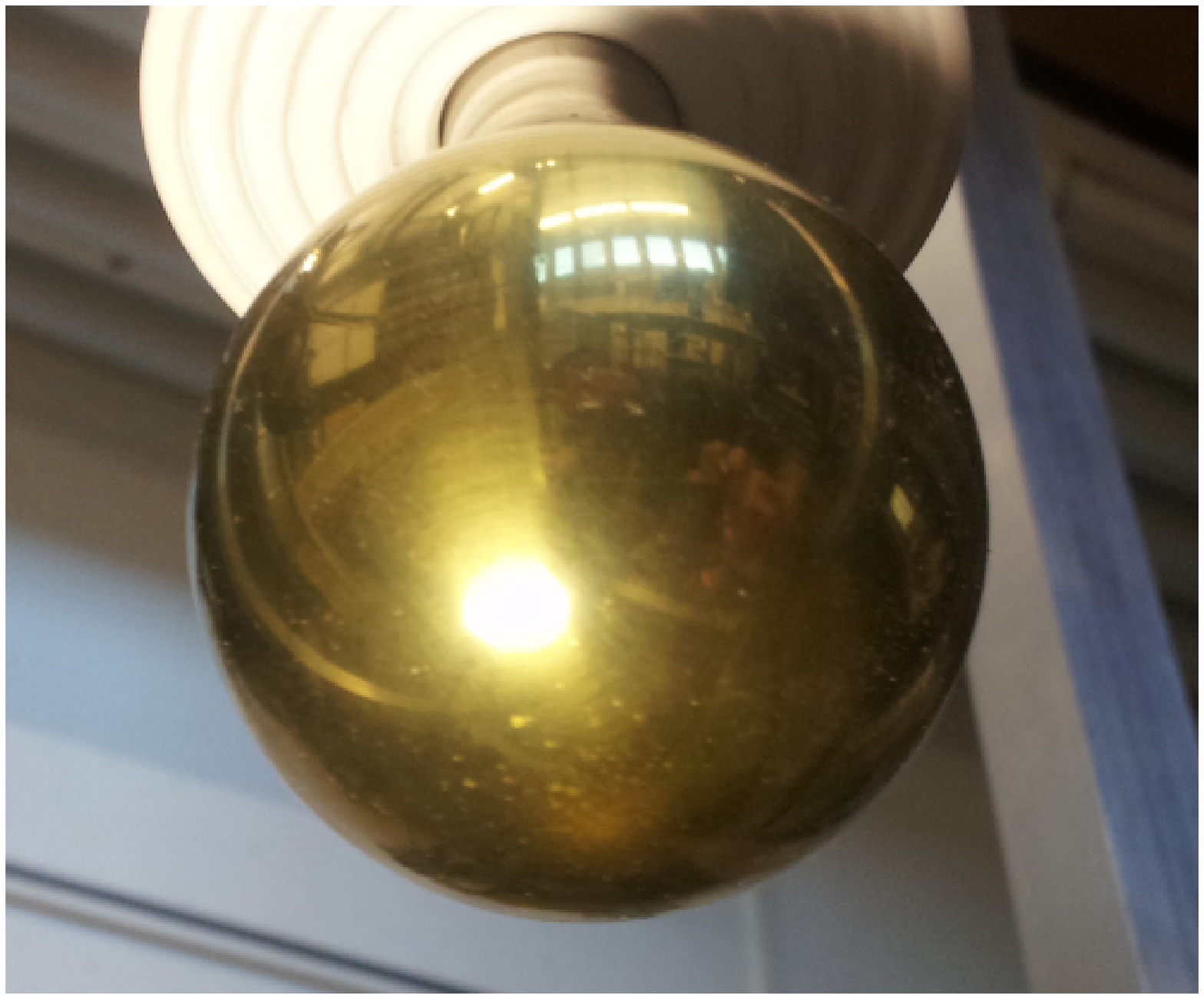}
\includegraphics[width=0.45\linewidth]{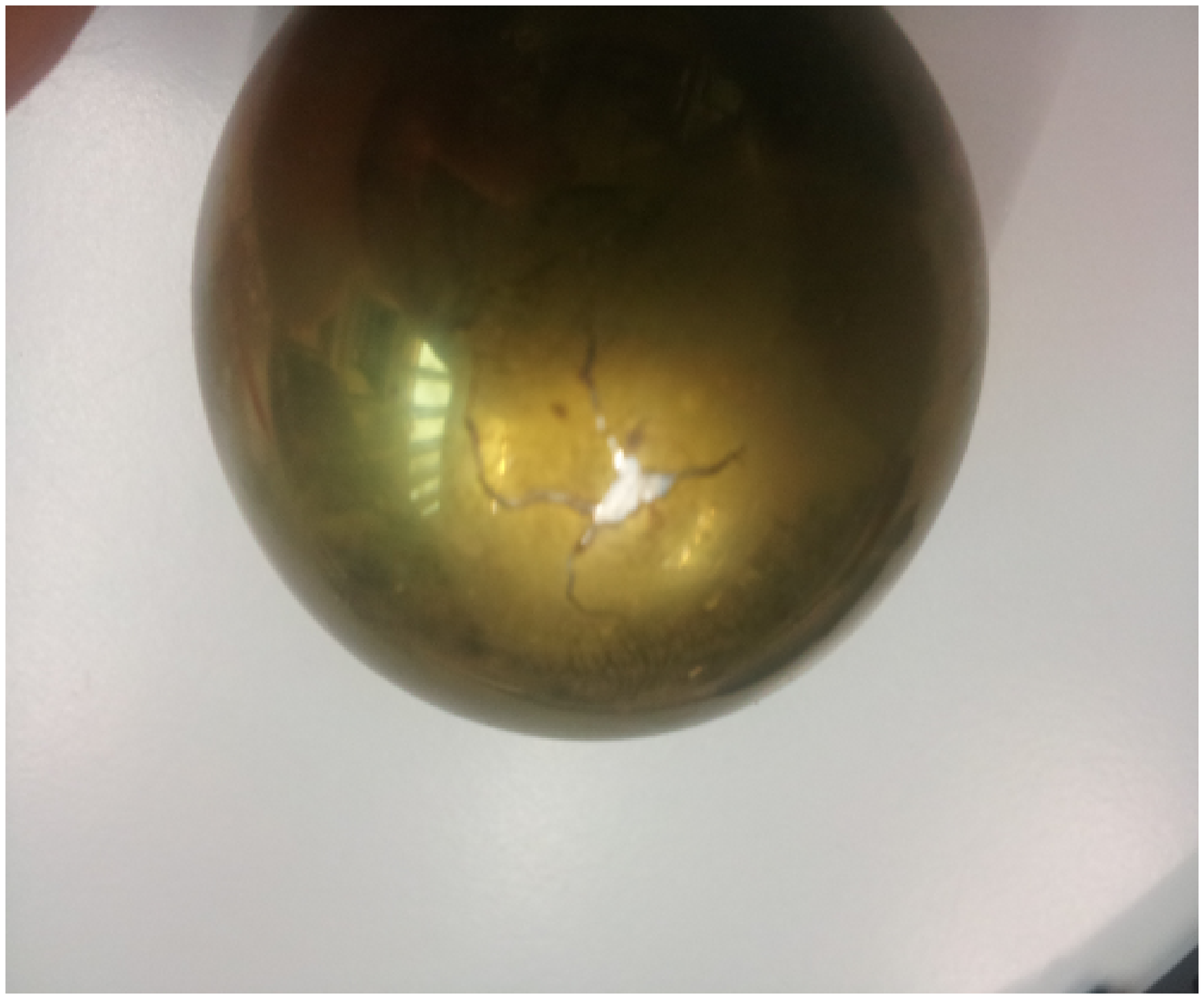}
\caption{Left: a 200 $\mu$m thick layer of natural polyisoprene deposited on the surface of a steel cathode sphere of 5 cm diameter. Right: layer damage after a discharge breakdown occurred.}
\label{fig:cathode}
\vspace{-3mm}
\end{figure}

A layer of polyisoprene deposited on the cathode sphere surface is shown in Figure~\ref{fig:cathode}, left. The deposition is made by dipping the sphere into purified latex milk, drying it at room temperature, leaching in deionized water for several hours and finally vulcanizing at 70$^{\circ}\rm$C for one hour. The leaching phase is crucial in order to remove all soluble pollutants contained in the natural latex milk. Vulcanization increases tear strength of the polymer layer.

\begin{figure}
\centering	
\includegraphics[width=0.45\linewidth]{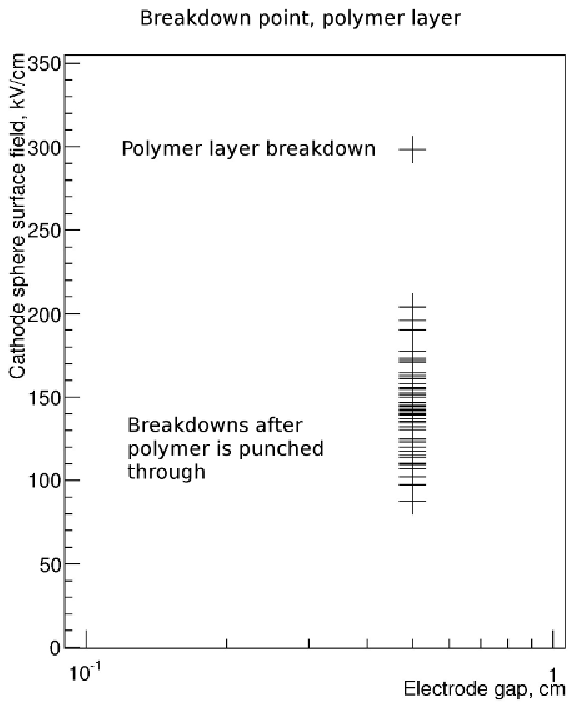}
\includegraphics[width=0.45\linewidth]{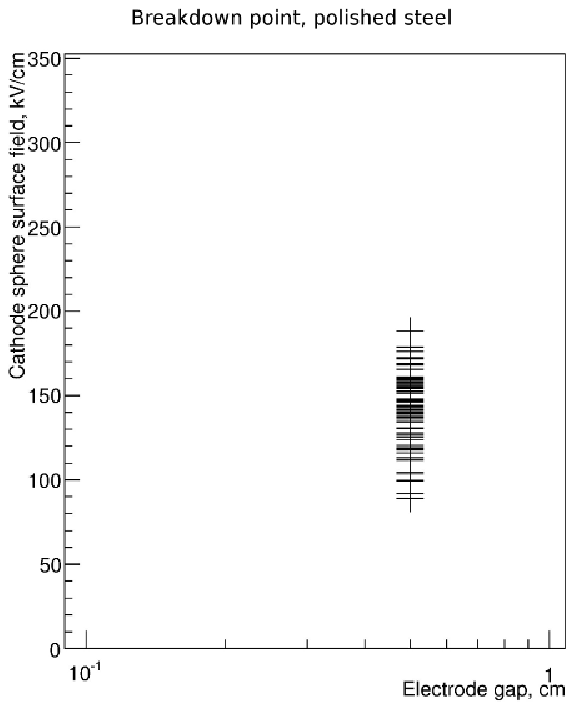}
\caption{Left: breakdown field intensity values in liquid argon for a cathode coated with a 200~$\mu$m thick polymer layer and a cathode sphere diameter of 5~cm. The top point corresponds to the first breakdown at 296 kV/cm, while the other points are due to breakdowns occurring when the layer has already been damaged. Right: for comparison, data are shown for uncoated steel cathode.}
\label{fig:both}
\vspace{-3mm}
\end{figure}

A first test was conducted with a 450~$\mu$m layer of polyisoprene deposited on the cathode sphere of 4~cm in diameter. After filling the cryostat with purified liquid argon (1~ppb of residual electronegative impurities), as described in \cite{Blatter:2014wua}, the variable cathode-anode gap was set to 5~mm. The voltage at the cathode was progressively increased at a rate of 50~V/s up to 130~kV, the maximum voltage delivered by the HV power supply. No breakdowns were observed during the following several hours in this steady regime. The voltage was then lowered to zero and the measurements were repeated with a gap of 4~mm, again, with no breakdown for several hours at 130~kV of maximum voltage.
After the gap was reduced to 3~mm the breakdown occurred during voltage ramp-up at the point of highest field (bottom of the sphere facing the anode plate) at 112~kV of cathode voltage, corresponding to a maximum field intensity of 412~kV/cm in the cathode-anode gap. This value is more than one order of magnitude higher than that at which we observed breakdowns in the absence of the polyisoprene layer \cite{Blatter:2014wua}.
The test was repeated with a 200~$\mu$m polyisoprene layer on a 5~cm diameter sphere. In this case, the breakdown occurred at a field intensity of 296~kV/cm for a 5~mm gap after about 1~min at 130~kV at the cathode.
After a breakdown occurs, the insulating layer is damaged (Figure \ref{fig:cathode}, right) and the breakdown field values lower to the level obtained with an uncoated steel surface, as reported in \cite{Blatter:2014wua} (see Figure \ref{fig:both}). This rules out any hypothetical effect of impurities emitted from the polyisoprene. Table \ref{table1} summarizes the results of the various measurements.

The mechanism responsible for the suppression of the cathode field emission of electrons is based on the very low mobility of excess electrons in the majority of solid organic polymers. Once the field at the metal surface rises to the value where field emission starts, the emitted electrons penetrate the insulating layer and slowly drift towards the interface with the liquid argon. The electron drift velocity in the polymer is a few orders of magnitude lower compared to that in liquid argon. Therefore, the secondary ionization is strongly suppressed in comparison to that occurring in liquid.
Electrons trapped in the polymer layer lower the surface field thus inhibiting further field emission. When they finally reach the interface, they help to neutralize argon ions that are created, for instance, by cosmic ray ionization. This, in turn, prevents ions from  accumulation at the liquid-polymer interface and inhibits the rise of the electric field across the polymer layer, which could possibly lead to an anticipated polymer breakdown. 

We expect that the solution we propose also works for cathode-ground distances larger than 1~cm, since the effect of the decreasing breakdown field intensity as a function of the distance that we reported in \cite{Blatter:2014wua} is closely related to development of field emission.

\begin{table}[ht]
\caption{Summary of the breakdown test measurements with 200 $\mu$m and 450 $\mu$m thick polyisoprene layers coating 5~cm and 4~cm diameter spheric cathodes, respectively.}
\vspace{1mm}
\centering	
\begin{tabular}{|l|l|l|l|l|}
\hline
Gap width & Max. field strength & Sphere diameter & Polyisoprene thickness & Breakdown \\ \hline
5 mm & 298 kV/cm & 4 cm         & 450 $\mu$m                 & no        \\ \hline
4 mm & 358 kV/cm & 4 cm         & 450 $\mu$m                 & no        \\ \hline
3 mm & 412 kV/cm & 4 cm         & 450 $\mu$m                 & yes       \\ \hline
5 mm & 296 kV/cm & 5 cm         & 200 $\mu$m                 & yes       \\ \hline
\end{tabular}
\label{table1}

\end{table}

\newpage
\section{Conclusions}

The deposition of a few hundred micrometer thick polyisoprene layer on the surface of a spheric stainless steel cathode immersed in liquid argon allows to efficiently suppress field emission of electrons from the cathode surface and to reach significantly higher electric fields intensities for cathode-ground distances of several millimeters. A field strength as high as 412 kV/cm was reached in the study reported in this paper. This solution allows to design and operate liquid argon TPC detectors keeping the volume of liquid  outside of the electron drift region comparatively much smaller than for the case of non coated cathode surfaces.



\end{document}